\documentclass[aps,prl,twocolumn,groupedaddress]{revtex4-1}
\usepackage{natbib}
\usepackage{url}
\usepackage{blindtext}
\usepackage{graphicx}
\usepackage{epstopdf}
\usepackage{epsfig}

\begin{document}


\title{The Secondary Ionization Wave and Characteristic Map of Surface Discharge Plasma in a Wide Time Scale}

\author{Yifei ZHU}
\email[]{yifei.zhu.plasma@gmail.com}

\author{Yun WU}
\email[]{wuyun1223@126.com}
\affiliation{$^1$ Science and Technology of Plasma Dynamics Laboratory, Xi'an 710038, People's Republic of China}
\affiliation{$^2$ Institute of Aero-engine, School of Mechanical Engineering, Xi’an Jiaotong University, Xi’an 710049, People's Republic of China}


\date{\today}

\begin{abstract}
Variation of voltage profiles with different time scales leads to the redistribution of deposited energy as well as electro-hydrodynamic forces, while the mechanism and scaling law is still unknown. On the basis of theoretical and numerical analysis, we show that a secondary surface ionization wave forms during the voltage rising slope when electron density decreases to a critical level while the voltage is still rising. A characteristic map of energy and electro-hydrodynamics force in time scales between 1~ns and 0.1~s at atmospheric pressure is proposed, opening the door towards the target--directed design of surface discharges.
\end{abstract}


\maketitle


The surface ionization wave above a dielectric has been the subject of many experimental and theoretical works since the 20th century. The discharge dynamics, the thrust and deposited energy are the key features of interest for different communities, e.g. surface treatment (some times called surface plasma ``bullets'')\cite{Lu2018}, atmospheric pressure plasma jet (APPJ), and plasma flow control by surface dielectric barrier discharges (SDBD)\cite{Jukes2009,Leonov2016}.
The advances of fast imaging techniques, modeling theories and tunable power sources in recent years have allowed deeper insights into the discharge dynamics and hydrodynamics characteristics of the surface ionization waves. The propagation of a surface ionization wave driven by a short nanosecond pulse is observed through optical spectrum emission and successfully reproduced by numerical simulation in \cite{zhu2017nanosecond,Babaeva2016}. A set of analytical solutions is built based on experiments and latest simulations~\cite{soloviev2015analytical}. In recent experiments~\cite{Zhang_2019,Huang_2020}, new power sources with tunable rising slopes are used, secondary current spikes are observed during the voltage rising edge, which are different from the widely known secondary current spike in the voltage trailing edge when the electric field is reversed. This phenomena has not been observed before and the mechanism is not clear yet. 

In this Letter we perform simulations to show the propagation of the surface ionization waves in conditions close to those of~\cite{Zhang_2019}, i.e., in air at atmospheric pressure driven by voltages of different rising slopes. The model results agree well with the experiments and can explain the observed secondary current spikes. They give a good quantitative prediction and explanation of the formation of the secondary ionization wave during voltage rising slope. Based on the experimental data, simulations and the analytical solutions summarized in this work, we propose a general scheme of energy and electro-hydrodynamic force characteristics of surface discharge plasma as functions of voltage rising time and amplitude.

\textit{The formation of the secondary ionization wave.} -- In the surface discharges, the secondary breakdown has long been observed at the trailing edge of the voltage pulse: the electric field and current change the direction at this stage, leading to the generation of a secondary surface ionization wave of the opposite polarity near the electrode.

At specific conditions, there exists a secondary ionization wave having the same polarity of charged head and current with the first one. This phenomena often occurs when the voltage (hundreds of kV) rising time is hundreds of microseconds and the gap is long ($\sim$1~m) and was known as the stepwise leader. The mechanism is closely related to the gas heating and changes in the gas density\cite{Bazelyan2020,Cheng2020}. 

However the secondary surface ionization wave occurs in a much shorter time scale (hundreds of ns) even before the first wave ends. We notice that this time scale corresponds to the electron decay due to the dissociative recombination process\cite{Soloviev2019}: 

\begin{equation}\label{eqs_electrondecay}
n_e(t)\approx\frac{n_{e0}}{1+t/\tau_r}
\end{equation}
\noindent
where $\tau_r^{-1}=1.4\times10^{-12}(300/T_e)^{0.5}n_e$ is the characteristic time of dissociative recombination of electrons with $\rm O_4^+$ ions, $n_{e0}$ is the electron density formed when the first ionization front passes, $t$ is the time after the first ionization wave passed.

When the electron density in the surface plasma channel decreases to $10^{18}-10^{19}~m^{-3}$, the in--channel region cannot shield the increasing external electric field. Thus, the high near electrode electric field plus the highly pre--ionized plasma trace left by the first ionization wave is possible to produce a secondary ionization wave. 

To formulate the critical electron density or time moment of the secondary surface ionization wave, the initial electron density $n_{e0}$ can be expressed as\cite{Soloviev2019}:

\begin{equation}\label{eqs_electrondensity}
n_{e0}\approx\frac{\nu_{ic}V_h^2\epsilon^2}{8\pi e\mu E_c^2d^2}
\end{equation}
\noindent
where the ionization frequency $\nu_i$ can be closely approximated by the dependence $\nu_i=\nu_{ic}(E/E_c)^2$, $\nu_{ic}=3.5\times 10^{11}~s^{-1}$ independent from gas density, and $E_c=277~kV/cm$ in atmospheric pressure. $\epsilon=4.3$ is the dielectric constant, $\mu\approx~0.06~m^2/(Vs)$ is the electron mobility. $V_h$ is the potential of the first ionization head and can be expressed as $V_h=V_{bd}+kt_{d}$, where $k$ is the voltage rising rate, $V_{bd}=7.3\times 10^4(1+1/(\pi\epsilon))h_d^{0.5}$ is the breakdown voltage, $t_d$ is the discharge propagation time:

\begin{equation}\label{eqs_td}
t_d\approx\frac{4 h_d^2\nu_{ic}}{\mu^2E_c^2\epsilon}
\end{equation}
\noindent
where $h_d$ is the thickness of the plasma channel ($\sim50~\mu m$ in atmospheric pressure). At critical time moment $t=t_{cri}$, the electron density decreases to $n_{ecri}$ when the secondary ionization wave occurs. Substituting equation~(\ref{eqs_electrondensity}), (\ref{eqs_td}) and $n_e(t)=n_{ecri}$ into (\ref{eqs_electrondecay}) leads to a quadratic equation. The solution plus the time before breakdown gives the criteria time moment of the second surface ionization wave:

\begin{equation}\label{eqs_criteria}
t_{cri} \approx 1.25\times 10^{12}\frac{n_{e0}-2n_{ecri}}{n_{ecri}^2}+\frac{V_{bd}}{k}
\end{equation}

The condition of secondary surface ionization wave formation can then be expressed as $t_{rise}\textgreater t_{cri}$ or $n_e\textless n_{ecri}$. It has to be noted that equation~(\ref{eqs_electrondecay}) is a rough approximation requiring the information of mean electron temperature $\rm T_e$. The accuracy of the estimation of $n_{e0}$ from equation~(\ref{eqs_electrondensity}) is affected strongly by the accuracy of $V_h$, $n_{e0}$ et al. To have a more precise calculation for equation~(\ref{eqs_criteria}), detailed numerical simulations are necessary.

We now take the experimental parameters as the input of our discharge model and investigate the condition of the secondary surface ionizationw wave formation. The discharge is modeled with the classical drift--diffusion--reaction model of Ref~\cite{zhu2017nanosecond}. Two voltage amplitudes are calculated: $V_{max}=14~\rm kV$ as in the experiment\cite{Zhang_2019} and $V_{max}=24~\rm kV$.

The height of the simulation domain is 5~cm and its weight is 8~cm, sufficiently much larger than surface ionization wave scale that the background field can be decided by Neumann boundary conditions for the electric potential on the boundary. Together with the dielectric with permittivity of 4.3 and thickness of 1~mm, this fixes the stationary field. The surface discharge plasma develops its own electric field. Because of the superposition principle, this can be calculated from the charge distribution within the discharge and then added to the other field.

The equations are discretized on a static nonuniform grid. The grid is refined in the area where a streamer is expected to propagate. The size of the finest grid cells is 5~$\mu$m. Away from the area of streamer propagation, grid cells exponentially increase in size up to 0.2 mm on the boundaries. To speed up the calculation, a smaller region with height of 0.1~cm and weight of 4~cm is used only to solve the drift--diffusion--reaction equations. 

\begin{figure}[t!h!]
	\epsfxsize=\columnwidth
	\begin{center}
		\epsfbox{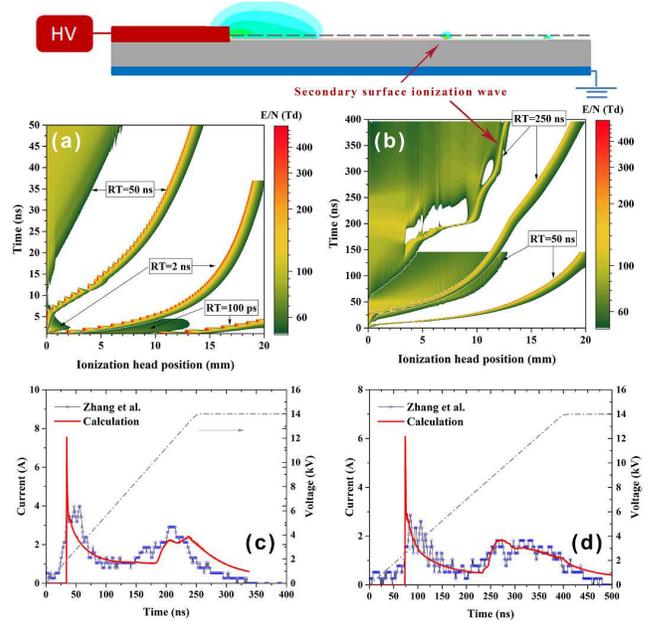}	
		\caption{The x-t diagram of the electric field in the discharge channel of 24~kV case at (a) $T_{rise}=100~ps, 2~ns$ and $50~ns$; (b) $T_{rise}=50~ns, 250~ns$; (c) the current value of the 14~kV case at (c) $T_{rise}=250~ns$ and (d) $T_{rise}=400~ns$. The dash dot grey lines in (c) and (d) are the applied voltages, the red solid lines are the calculated discharge current and the scatters are from experimental measurements.\label{figExperiment}}
	\end{center}
\end{figure}


The result of our simulations is shown in Fig.~\ref{figExperiment}. Fig.~\ref{figExperiment}~(a-b) show the x-t diagram of the field in the discharge channel (25~$\mu m$ above the dielectric) at 24~kV with different rising times ($t_{rise}$=100~ps, 2~ns, 50~ns and 250~ns). For the same peak voltage, the ionization front propagates much faster with higher electric field at shorter rising times (the left panel). With the increase of the voltage rising time, the field near the electrode grows and a secondary ionization front appears starting from 150~ns in the $t_{rise}$=250~ns case. 

The calculated current values for the 14~kV case with $t_{rise}$=250~ns and $t_{rise}$=500~ns are plotted in Fig.~\ref{figExperiment}~(c-d) together with experimental measurements and applied voltage profiles. The starting moment and the amplitude of the secondary current peaks agree well with the analytical prediction and experimental measurements.

We take the electron density value from the discharge model as the input for equation (\ref{eqs_criteria}) to calculate the criteria electron density $n_{ecri}$. At $V_{max}=24~\rm kV$ and $t_{rise}=250~ns$ condition, $n_{e0}=1.6\times 10^{20}~m^{-3}$ and $t_{cri}=150~ns$, this leads to $n_{ecri}=5.1\times 10^{19}~m^{-3}$. At $V_{max}=14~\rm kV$ and $t_{rise}=250~ns$ condition we have $n_{ecri}=4\times 10^{19}~m^{-3}$. The same procedure applied to $V_{max}=14~\rm kV$ and $t_{rise}=400~ns$ condition leads to $n_{ecri}=3.7\times 10^{19}~m^{-3}$. Thus, if the voltage rising time is longer than the time required for the electron density decaying to below $3\sim5\times 10^{19}~m^{-3}$, it is possible that the secondary surface ionization wave appears.

\textit{The energy deposition.} -- Energy deposition is the pioneering process of gas heating/fluid responses and has been found to be related with the rising slope~\cite{Benard2012}. In general, shorter rising time leads to higher energy deposition (calculated from $j\cdot E$), but quantitative analysis is not available. 

\begin{figure}[t!h!]
	\epsfxsize=\columnwidth
	\begin{center}
		\epsfbox{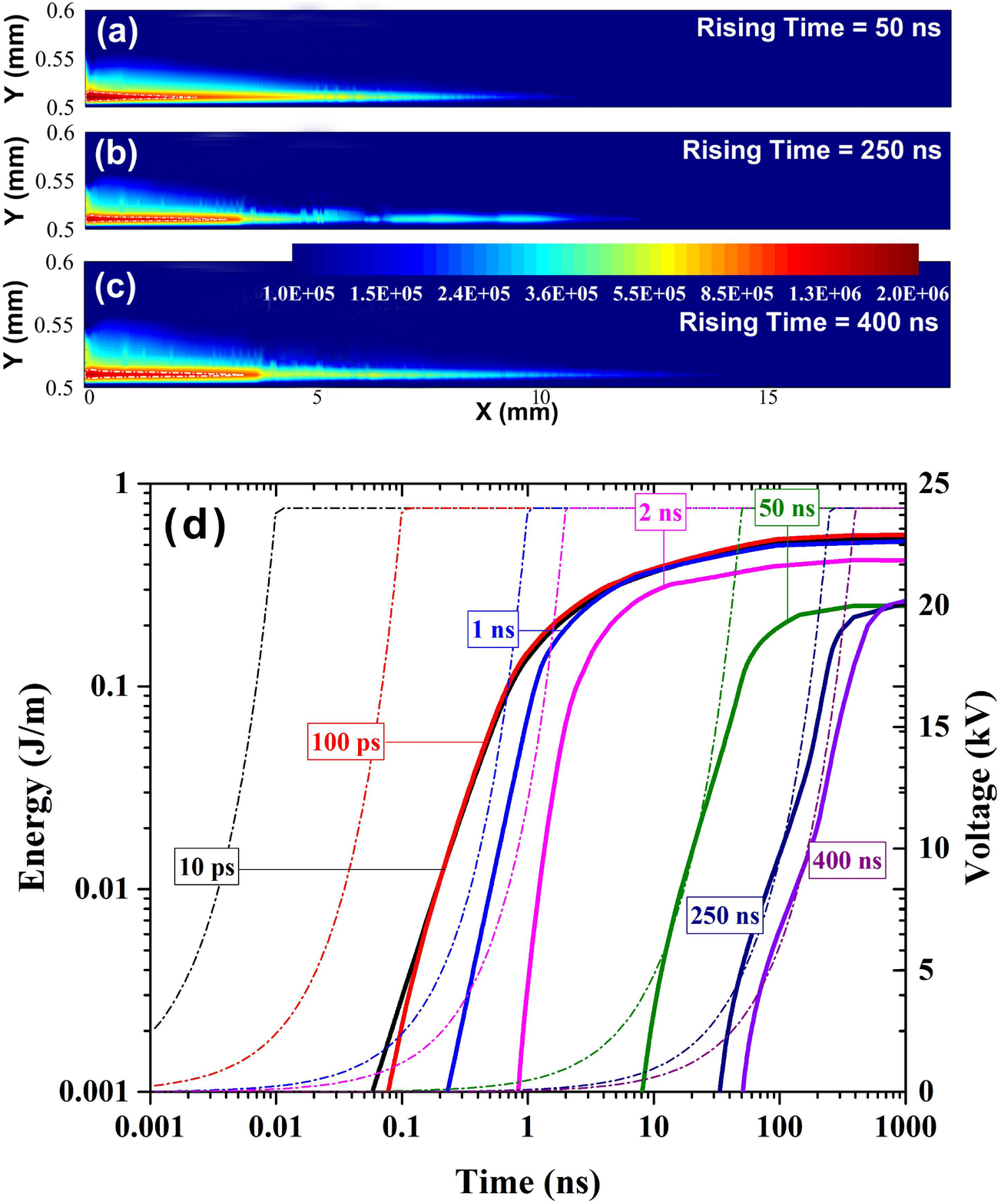}	
		\caption {The spatial distribution of the total deposited energy at different rising times (a) $t_{rise}=10~ps$ (b) $t_{rise}=250~ns$ (c) $t_{rise}=400~ns$ and (d) the temporal evolution of total energy. The dash--dot lines are the applied voltage and the solid lines indicate the energy deposition from 1~$ps$ to 1~$\mu s$.}
		\label{figExperiment2}
	\end{center}
\end{figure}

We calculate the total energy deposition for different rising slopes shown in Fig.~\ref{figExperiment2}. The upper panel shows the spatial distribution of deposited energy. The energy is distributed smoothly along the plasma channel at short pulses (Fig.~\ref{figExperiment2}~(a) and \cite{zhu2018fgh}). At longer rising time, the secondary ionization wave appears, and the energy is also deposited in the entire channel but are mainly in a local region behind the secondary ionization wave because of higher electric field as predicted in Fig.~\ref{figExperiment}.

The lower panel of Fig.~\ref{figExperiment2} shows the temporal evolution at different rising slopes. It is interesting to find that, with a fixed peak voltage, the deposited total energy increase with the shorter pulse front until $t_{rise}= 1~ns$, and decrease with longer pulse rising time. In the moment of the secondary ionization wave appearance (T$\le$250$\mu s$), the energy deposition is more or less the same. This phenomena can be explained starting from the viewpoint of an analytical solution. 

The total deposited energy of SDBD was formulated in~\cite{soloviev2015analytical} as a function of permittivity, voltage, discharge length, dielectric thickness and breakdown voltage:

\begin{equation}\label{eqs_energyformula}
Q\approx\frac{\epsilon V_{max}^2l_s}{16\pi d}(1-\frac{V_{bd}^2}{V_{max}^2})\times 10^{-9}
\end{equation}
\noindent
where $l_s$ is the discharge propagation length. In a pulse discharge, $l_s$ is the discharge propagation length, while in SDBD driven by sinusoidal voltage waveforms, $l_s$ is the accumulative propagation length due to successive micro--discharges\cite{soloviev2015analytical,Soloviev2019}.

We rewrite equation~(\ref{eqs_energyformula}) to take into account the different morphologies of the discharge (3D filaments or 2D sheets) and the voltage rising slopes in following conditions:

(1) $t_{rise}\textless t_d$: in this case the voltage rising slope is ultrashort (a few nanoseconds), the discharge is in the form a quasi--uniform 2D discharge sheet. In this case we consider the discharge as a single 2D sheet, $l_{s}=2 h_d\nu_{ic} V_{max}/(\mu {E_c}^2)$.

(2) $t_{rise}\gg t_d$: we consider $t_{rise}\textgreater 10t_d$, in this case the voltage rising slope is slow, there can be multiple discharges appear simultaneously in the form of 3D filaments, in this case $l_{s}=l_{s2D}(5\epsilon+4)log(1+2d/h_d)/(6(\epsilon+1)(\epsilon+2))$, is the propagation distance of a single discharge layer of separated 3D filaments.

(3) $t_d \textless t_{rise}\textless 10t_d$: it is a transition stage when the rising time and discharge propagation time are comparable. In this case, both conditions are possible: there is only one discharge before the maximum voltage in the style of a 2D sheet, or just like in the case of condition (1), it is hard to have a definite formula for this condition, thus we assume the energy are decreasing linearly with the increasing $t_{rise}$.

\begin{figure}[t!h!]
	\epsfxsize=\columnwidth
	\begin{center}
		\epsfbox{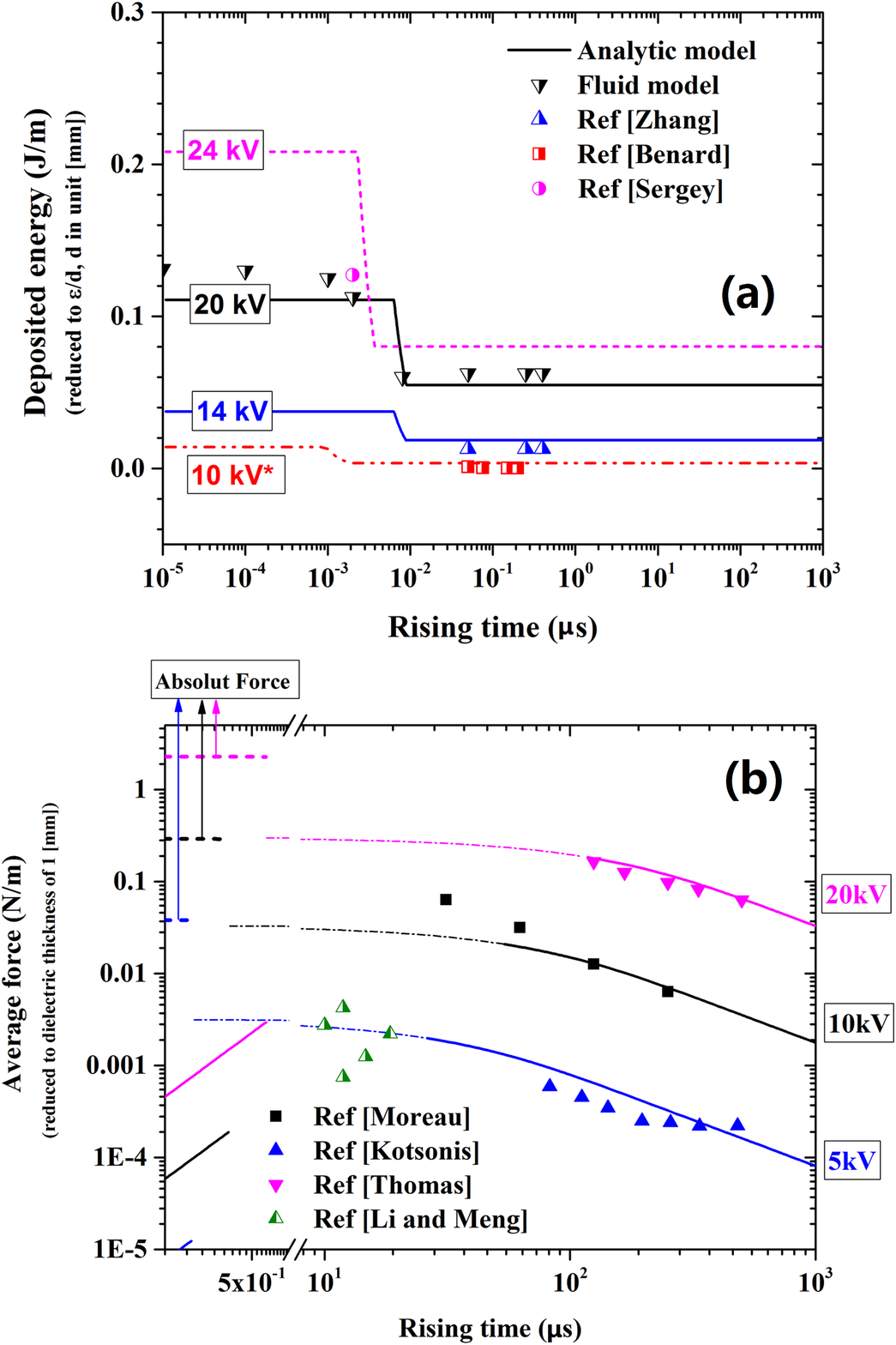}	
		\caption {(a) The deposited energy (reduced by dielectric permittivity and thickness) as a function of $t_{rise}$. The solid, dash and dash--dot lines are theoretical results (different line styles indicate different $\epsilon/d$ values). The symbols correspond to numerical simulation, and experimental data extracted from Ref~\cite{Zhang_2019,Benard2012,Shcherbanev2016}. The 10~kV case marked with a star is special, the total deposited energy is quite small, but if we plot this figure in log Y scale, the deposited energy of the 10~kV case is not a constant.  (b) The average force as a function of $t_{rise}$. Solid lines are the thrust value averaged to one duty cycle, the dash--dot lines extended from solid lines are calculated results in the ``saturated'' region and only for reference. The dash lines are the thrust in one pulse. The experimental data are extracted from paper~\cite{Debien2012,Kotsonis2011,Thomas2009,Meng2013}. }
		\label{figExperiment3}
	\end{center}
\end{figure}

The results of the theoretical estimation of total deposited energy are plot in Fig.~\ref{figExperiment3}~(a) with measurements from independent groups of experiment and simulations. It is clearly seen that there is a significant rise of deposited energy when $t_{rise}$ is reduced to the value comparable to $t_d$, agreeing well with both simulations and experiments.

\textit{The electro--hydrodynamic force generation.} -- In a surface discharge, the electro--hydrodynamic force is caused mainly by the motion of negative charged particles $\rm O^-$ and $\rm O_2^-$. An analytical expression of the time averaged thrust generated by a sinusoidal ac voltage driven surface discharge has been formulated in \cite{soloviev2015analytical}, which considers the accumulative charge by the micro--discharges series occur in one sinusoidal period, and applies only to the sinusoidal voltage waveform. When the voltage rising time is decreased to the streamer propagation time scale (tens of nanoseconds), the successive mirco--discharges disappears and an uniform discharge happens. We rewrite above formula as a function of $t_{rise}$ to make it more general. Two parameters are defined here for following discussions, the time interval between micro--discharges $t_{md}=180[{\rm V}]\times t_{rise}/V_{max}$ and the relaxation time of the negative charge in the surface plasma channel $t_{res}=l_s/V_i$ ($V_i\approx 100~\rm m/s$ is the ion drift velocity).

(1) $t_{md}\textless t_d$, the time interval between micro--discharges is smaller than the discharge propagation time, that is, the voltage rising time is short enough that only one discharge occurs, this is often the case of nanosecond discharges. Assuming the pulse duration time is two times the voltage rising time, the average force in one pulse $F_{pulse}$ is:

\begin{equation}\label{eqs_thrustformula1}
F_{pulse}=2.4\times\frac{1}{t_{rise}d}(1-exp(-\frac{2t_{rise}}{t_{res}}))l_{s2D}^4
\end{equation}

(2) $t_{res} \textless t_{rise}$: the voltage rising time is long enough for the negative charge to be relaxed in the channel, thus successive micro--discharges occur in one duty cycle, this is often the case of sinusoidal SDBDs. The average force in one period $F_{pulse}$ is:

\begin{equation}\label{eqs_thrustformula2}
F_{ac}=2.4\times\frac{1}{t_{rise}d}(1-exp(-\frac{t_{rise}}{t_{res}}))l_s^4
\end{equation}
\noindent
where $l_s=1.5\times10^{-4}(1+(9V_{max}/4dV_c)(1-7dV_c/6V_{max}))$, $dV_c=600~[\rm V]$ is the normal cathode voltage fall. If $t_{res} \textgreater t_{rise}$, the discharge is saturated, negative charges cannot be fully relaxed and the discharge turn to be filamentary, equation~(\ref{eqs_thrustformula2}) may fail. 

The calculated electro--hydrodynamic force is shown together with the measurement by different groups worldwide in Fig.~\ref{figExperiment3}~(b). The solid lines are the calculated period average thrust, the dash-dot lines mark the region where the discharge become saturated, while the dash lines indicate the pulse averaged thrust. The calculation and the experiment are in good agreement.

\textit{The characteristic map.} -- The combination of existing experimental results, the numerical simulations and the analytical theory gives an excellent description of the behaviors of surface ionization waves in atmospheric pressure, and allow us to draw a general two--dimensional map characterizing the reduced Electro-Hydrodynamics force and total energy deposition as a function of voltage amplitude and rising time. 

\begin{figure}[t!h!]
	\epsfxsize=\columnwidth
	\begin{center}
	\epsfbox{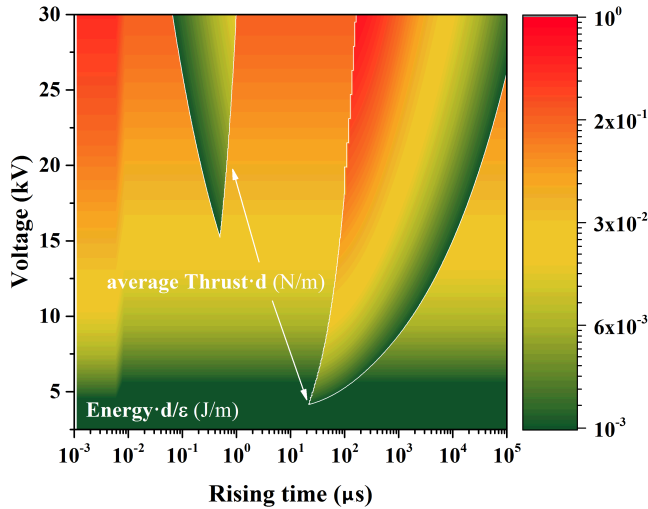}	
	\caption {The characteristic map of total deposited energy and thrust in a surface dielectric barrier discharge in atmospheric pressure with $t_{rise}$ ranging from 1~ns to 0.1~s. The maps of deposited energy and thrust are overlapped, the energy (reduced by dielectric thickness in mm and permittivity) map is the background and the two ``watermelon''--like petals are the map of average thrsut (reduced by dielectric thickness in mm).}
		\label{figExperiment4}
	\end{center}
\end{figure}

We plot the characteristic map in Fig.\ref{figExperiment4}. The map is an overlapping of the thrust map (the two ``watermelon''--like petals with the yellow outlines) and the energy map (the background). Increasing the voltage, in general, leads to higher thrust and energy deposition. When the voltage rising time is reduced to the scale of surface ionization wave propagation time (a few to tens of nanoseconds), the deposited energy will increase significantly. To achieve higher average thrust, one has to move to the red region of the ``watermelon''--like petals, which is also the ``sweetest'' region in a real watermelon.

\textit{Summary.} -- We have found that the requirements of the formation of the secondary surface ionization wave is the decay of electron density below $3\sim5\times10^{19}~m^{-3}$ while the voltage pulse is still rising. The reason is that the field in the streamer channel behind the ionization head cannot shield the still rising external electric field originates from the electrode. An analytical expression is proposed to decide the transition moment. A 2D model coupling Poisson’s equation and plasma drift--diffusion equations at atmospheric pressure can reproduce strikingly well the features of secondary surface ionization wave observed during discharge breakdown driven by sub--nanosecond pulses at atmospheric pressure. 

The energy deposition and electro--hydrodynamic force are studied combining experiments, the 2D model and an extended analytical model. Calculations performed for different voltage rising times and amplitude show good agreement with the experimental results. The secondary surface ionization wave results in the spatial redistribution of the deposited energy but does not affect the total energy deposition. Decreasing voltage rising time below the characteristic discharge propagation time leads to a sharp increase of total energy deposition. 

The combination of experiments, numerical simulations and analytical theories allow us a deeper insight into the physics of surface ionization waves. A general characteristic map was drawn, providing a direct and clear description of the performance of surface ionization waves, this map links the theoretical work and applications, and opens the door towards the target-directed design of the surface discharges.

The work was partially supported by the National Natural Science Foundation of China (No. 51907204, 51790511, 91941105, 91941301) and the National Numerical Windtunnel Project NNW2018-ZT3B08. The authors are thankful to the young research group in Atelier des Plasmas for fruitful discussions.

\bibliography{Libaray_PRLSDBD}
\bibliographystyle{apsrev4-2}
\end{document}